\documentclass[multphys,vecphys]{promult}

\usepackage{makeidx}     
\usepackage{graphicx}    
\usepackage{multicol}    

\makeindex             

\begin{document}

\title*{RTS2 -- the Remote Telescope System}
\author{Petr Kub\'anek \inst{1,2}}
\institute{Image Processing Laboratory (IPL), Universidad de Valencia, Valencia, Spain
\and IAA CSIC Granada, Spain}
\maketitle

{\it RTS2} is an open source observatory manager. It was written from scratch
in the C++ language, with portability and modularity in mind. Its driving
requirements originated from quick follow-ups of {\it Gamma Ray Bursts}. After
some years of development it is now used to carry tasks it was originally not
intended to carry. This article presents the current development status of the {\it RTS2}
code. It focuses on describing strategies which worked as well as things which
failed to deliver expected results.

{\it \small Copyright 2010 Petr Kub\'anek. This is an open access article distributed under the Creative Commons Attribution License, which permits unrestricted use, distribution, and reproduction in any medium, provided the original work is properly cited.}

\section{Introduction}
\label{sec:1}
{\it RTS2} originates from {\it RTS}, {\it the Remote Telescope System}. {\it
RTS} was written during years 1999 and 2000 as a team project for Computer
Science courses\cite{jilek00} by students of Mathematics and Physics faculty of
Charles University in Prague.

{\it RTS2} is primarily developed on the {\it Ubuntu Linux} distribution. It is
known to run on a wide variety of Linux distributions, the {\it Solaris} operating
system and partially on {\it Microsoft Windows} and {\it Mac OS X}.

Project source code can be obtained from the Web site {\tt http://rts2.org}. The project Wiki
pages on {\tt http://rts2.org/wiki} list group experiences with various fully
autonomous observatory projects.

The outline of this paper is as follows: The first section is this
introduction. The next section describes goals and history of the project
development. The third section focus is a description of how various tasks inside
{\it RTS2} are divided.  The fourth summarises communication protocol. The next section
describes approaches used during development. The sixth section forms the core of the
article -- it lists currently added features. The seventh section informs readers
about currently running {\it RTS2} systems. The eighth is a reminder about our
experiences with restarting failed devices. The contribution concludes with
a section about expected developments.

\section{Project goals and its development}
\label{sec:2}

The original goal of the {\it RTS2} development was to produce software for a
small robotic telescope, devoted to study Gamma Ray Burst ({\it GRB})
transients. The requirements of the project were:

\begin{itemize}

\item must be fully autonomous

\item must react as fast as possible to incoming GRB alerts

\item must perform regular observations during periods with no GRB observation to verify system readiness

\item must be modular to enable easy switching of the instruments

\end{itemize}

Those requirements were laid in early 2000. They share similarities with
other projects, notably {\it AudeLA}\cite{klotz08} and the {\it Liverpool
Telescope System}\cite{fraser02}. At the time, open--source projects of similar
scale did not exist. Available commercial, closed--source solutions were either
associated with a single instrument, with questionable portability to others
telescopes, or did not fulfill the requirement for fast switching between
targets. It is worth noting that even today it is not easy to implement this
requirement. It forms a strong entry barrier for those who would like to become
{\it GRB} observers.

As the project advanced, additional requirements were added. They reflected
experiences gained during development, and particularly pains and problems
associated with early porting of the system to the other observatories:

\begin{itemize}

\item system must be robust enough to continue operations even when
non--critical part(s) fail

\item observer must have possibility to remotely interact with observations

\item system must be fully configurable using configuration files

\item system must provide clear description of what its components are performing

\item observer must be presented with a list of devices which failed

\item the code should include dummy device drivers for testing the software without hardware

\end{itemize}

Creation of dummy environments brings great benefits during system
development. They enable developers to debug the system before it is deployed.
There are still cases when errors and bugs are detected only during night runs.
But as the project matures, the number of those cases significantly decreases --
and there are recorded cases when new software, with significant new features,
was deployed and it just worked, without any debugging.

As the number of observatories running {\it RTS2} grows, their management,
fixing various small glitches, as well as their scheduling and data processing
started to saturate staff time. We are aware of this, and we are in progress of
creating tools which will allow us to manage network operations using a smaller
amount of operators' time.

\section{RTS2 quick overview}
\label{sec:3}

{\it RTS2} is based on the {\it "plug and play"} philosophy. The parts which
constitute the system can be started, restarted or stopped anytime, without
affecting system performance. Special care is taken of resolving all possible
blocking states, so the system will always respond to requests in reasonable
time.

The whole system is user--space based and except for drivers provided either by
hardware manufacturers or by our group, it does not require any kernel--space based
components.

Code is designed around a central {\it select} system call, which picks any
incoming messages. If there is no incoming message, class {\it idle}
method is called. The code uses extensive hierarchy of its own C++ classes.

\begin{figure}
\centering
\includegraphics[width=12cm]{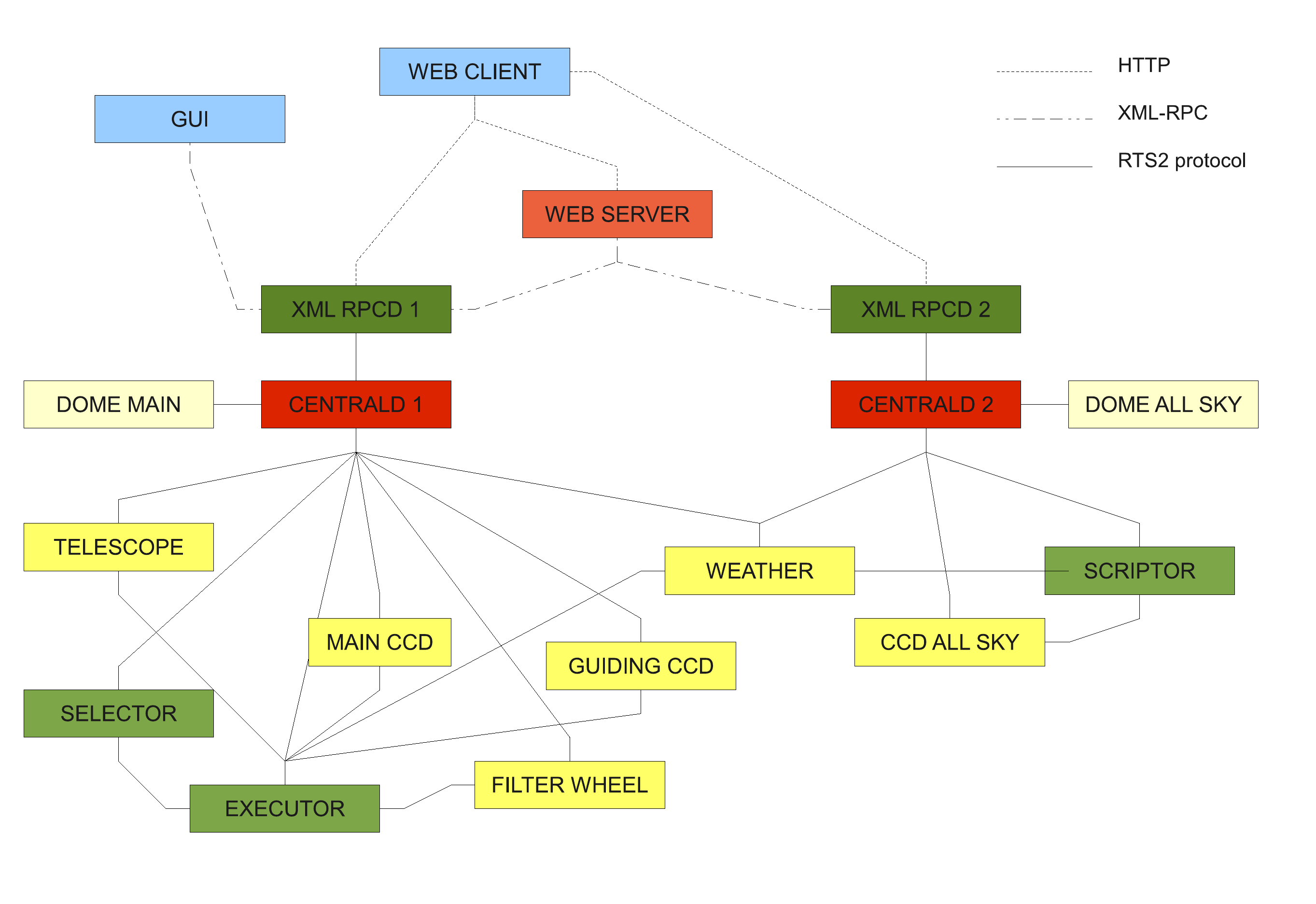}
\caption{Example RTS2 environment. Two basic setups are present on the site --
a telescope with a guiding CCD and a main CCD equipped with a filter wheel, and an
all sky camera. There are two domes (one for telescope, second for all sky
camera), three CCD detectors, {\it executor} and {\it selector} services
controlling telescope setup and {\it scriptor}, a simplified executor service,
controlling the all sky camera.  Both observatories offers external access
through XML-RPC protocol, provided by {\it XMLRPCD}. This is used by the Graphical
User Interface and the Web server.  {\it XMLRPCD} also provides a Web browser with
direct access to some functions. Access to XML-RPC and Web functions can be
protected with a password.}
\label{fig:environment}
\end{figure}

The {\it RTS2} system consists of the processes summarised in table
\ref{rts2quick}. Figure \ref{fig:environment} shows processes and connections used on
an example observatory, which includes three CCDs, two domes and a single
weather station serving both domes. For a detailed description please see
\cite{kubanek06}.

\begin{table}
\centering
\caption{RTS2 processes}
\label{rts2quick}
\begin{tabular}{l|p{10.5cm}}
{\bf Process}         & {\bf Description} \\
\hline
\hline
centrald  & Central component of the {\it RTS2} system. It provides three main services -- a list of devices and services present in the system, system state changes and synchronisation among devices \\
\hline
devices   & Corresponds to hardware attached to the observatory system. Different classes of devices are provided. They represents hardware coming from different manufacturers. \\
\hline
services  & Represents execution logic of the system. They provides functions for the end user - carry observations, receive GCN and other alerts, enables XML-RPC\footnote{XML-RPC web site: {\tt http://www.xmlrpc.com}} based access to the system. \\
\hline
clients   & Provides information to end--user. They are usually run in interactive mode, with end--user interacting with the programm. They include an interactive monitor and simple tools to execute observation scripts. \\
\end{tabular}
\end{table}

\section{Communication protocol}

{\it RTS2} employs its own communication protocol. A detailed protocol description
can be found in \cite{pkub08}. The protocol is based on sending ASCII strings over
TCP/IP sockets. It is fast, simple and robust. Among its important features are:

\begin{itemize}

\item sending of {\it "I am alive traffic"}, disconnecting connection if reply
is not received - this removes connections to dead components

\item ability to switch to binary mode for data transfer, to transfer images
and other large data items

\item traffic speed--up -- values which were not changed are not transported

\end{itemize}

The protocol supporting libraries provides developers with a flexible and
simple way to use common parts of the system. This results in a robust, 
transparent code, which can be easily extended. This greatly enhances software
reuse. There are reasonable hopes and claims that the interfaces are simple
enough to be understand by developers unfamiliar with the project.

\section{Coding style \& philosophy}

It is widely rumoured that there are as many approaches to coding as there are
software developers. This section aims to list rules and customs used during
{\it RTS2} development. Although most of them are widely known we hope they are
worth mentioning.

The best documentation for the code is the code itself. Readers of this article
are encouraged to check out the system and study it. Code with complexity of the
{\it RTS2} project cannot be made completely transparent. It is not expected that
the developer will understood code on the first encounter. It is important that
he/she find interfaces being used. Later he/she can continue progressing
towards system core classes.

\subsubsection{Track software changes in Version Control System}

The project is tracked in Subversion\footnote{Subversion web site - {\tt
http://subversion.tigris.org/}} tracking system. Previously version tracking
relied on Concurrent Version System (CVS)\footnote{CVS web site - {\tt
http://www.nongnu.org/cvs/}}. We have to admit that switch of the version
control system to Subversion provides significant improvements, and we were
really surprised with Subversion capabilities. Subversion really provides
simply resolution for situations which were difficult or impossible to handle
with CVS.

\subsubsection{Code releases}

As the project is still under active development, a release formalism is yet not well
established. There are usually two big releases during a year. All
observatories are running Subversion code -- code is regularly updated from
Subversion and put to use. This way version management is also used for
software distribution. We expect to calm this pace and establish a formal
release mechanism.

\subsubsection{Development cycle -- release early}

Changes are usually committed to Subversion as soon as they compile without any
errors. Before each commit, difference between new code and code in repository
is reviewed. Sense and purpose of committed lines is examined once again.

This approach tracks vast majority of bugs right before they make it to the
version control repository. After all changes are committed, and at least some
documentation which explains new features is provided, behaviour of the code is
tested once again. It is up to developer, where tests will be performed.
Usually the system is tested on dummy devices, before moving to a real observatory.
But small changes, which should introduce predictable results, are tested
directly on observatories.

\subsubsection{Follow common practices}

The common practices can be summarized by a few sentences, such as: Think
twice, code once. Discuss with others, inform users about changes.  Try to
design a generic solution instead of a simple additions for new problems.  Add
new features slowly and test them before implementing another.  Divide a complex
problem to simpler subproblems, implement and test the solution for them first,
and then integrate the code to solve the complex problem. Keep in mind that the
best developers are able to design, code, test, and release no more then 100
lines per working day -- try to keep number of lines small by reusing what is
already available either in your code or in C++ libraries.

\section{Current developments}
\label{sec:3}

This section deals with features which were recently added to {\it RTS2}. The
list provided bellow is not complete -- please see project change--logs for
a more in--depth description.

\subsection{State machines}

{\it RTS2} uses state machines. The states represents various states of the
hardware -- for example camera can have idle, exposing and readout state.

The states were originally used for coding purposes, to distinguish various
states of the code. They were displayed in monitoring applications, so the user
was informed what the device should do.

\begin{figure}
\centering
\includegraphics[width=12cm]{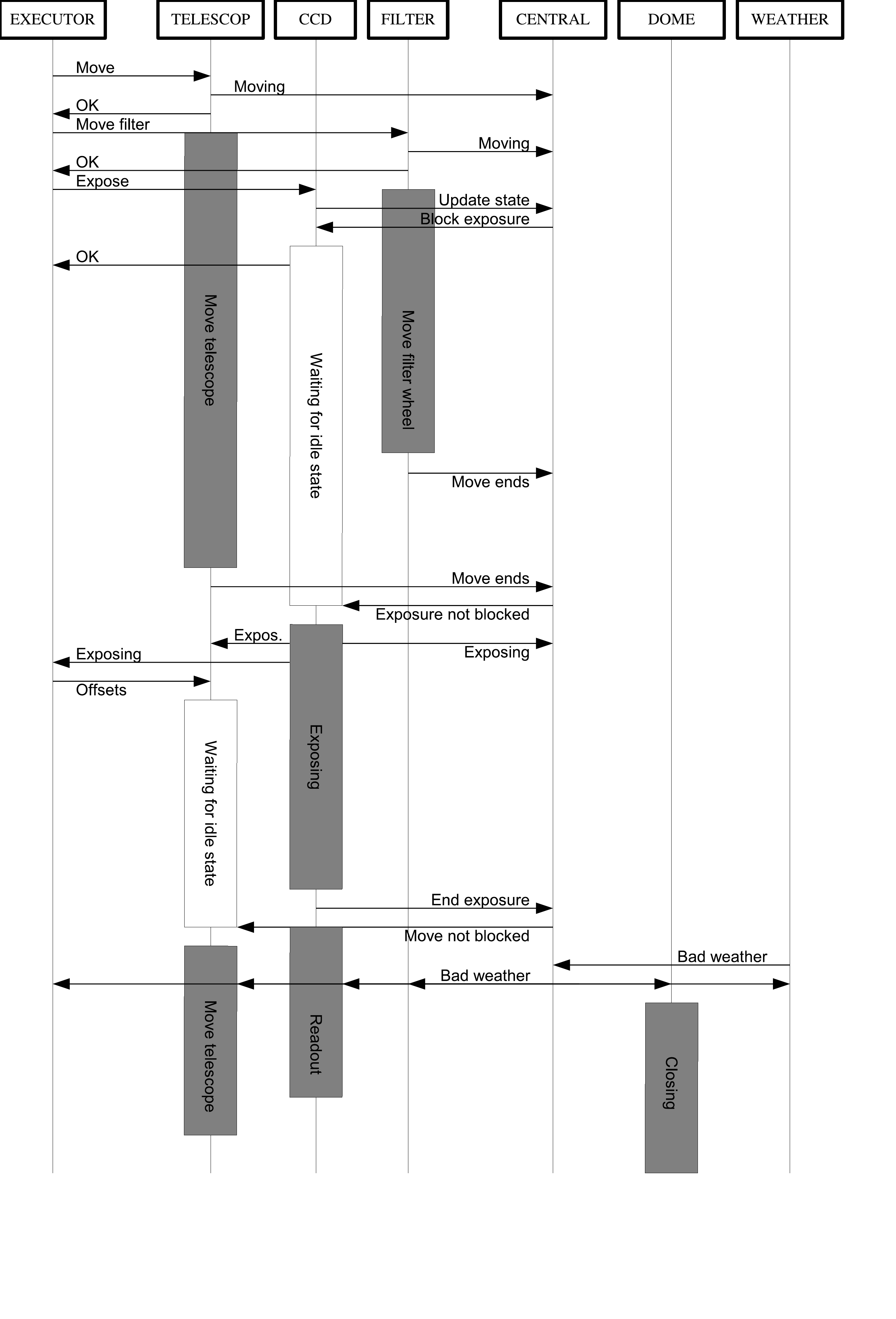}
\caption{Interactions of devices using states for synchronisation. Note how
various block are placed to keep telescope from moving and camera from
exposing. The observation is interrupted by bad weather reported from a weather
sensor, which closes the roof.}
\label{swiming}
\end{figure}

Later state use was expanded toward synchronisations. States prevent the camera
from taking exposures during telescope movement, and unwanted telescope movements
during camera exposures. They are displayed on users displays, allowing
observes to identify which device is blocking the next exposure or the next telescope
movement. The use of states for synchronisation is depicted in figure~\ref{swiming}.

\subsection{Default configuration}

Originally {\it RTS2} did not change configuration of devices prior to
observations. That leaves the user responsible for device configuration.  He/She
can do that using either a monitor application or a script for observation.

This handling was later extended by {\it RTS2} remembering default values. The
command {\it script\_ends} was added to the protocol. It loads back all changed
values.

This code was successfully used for focusing. The system was able to change
back to the last known good focuser position even when a focusing run was interrupted.
But it starts to show its bottlenecks, notably:

\begin{itemize}

\item unpredictable behaviour for the end user -- he/she changes something, and
suddenly another value appears in acquired images

\item unknown default values -- system unfortunately does not show which values
were changed

\item the system was quite hard to code, and as a consequence, quite hard to
maintain

\end{itemize}

Various solutions were considered to resolve those bottlenecks. The following
was selected as the best one:

\begin{itemize}

\item when appropriate, the system offers default value and various offsets.
The default value can be changed only on user request, and is never changed by the
system. Offsets can be changed by user and/or system. Offsets are zeroed before
beginning of the next observation.

\item or default values are provided only for values written in devices default
configuration file.

\end{itemize}

Both cases can be best understood from an example. The first case can be best
demonstrated on the new focuser interface and the second on handling mode settings
on CCD.

\subsubsection{Offsets, default and target values}

The interface provides end user with the following values:

\begin{itemize}

\item {\bf current value}, which is read from focuser

\item {\bf target value}, equal to sum of values listed below

\item {\bf default value}, set by the user

\item {\bf focusing offset value}, settable by the user

\item {\bf temporary focusing offset value}, settable by the user, zeroed at the end of the observing script

\end{itemize}

Changes of the last three values are used for user focuser interaction.
The default value is changed when the user would like to significantly change focuser
position. For example the offset can be set for different optical thickness of
the filters. And temporal focusing offset is used in the focusing script to change
the focuser position during focusing runs.

If a focusing run is interrupted, the focusing script finished, or the focusing program
is disconnected, the temporal focusing offset is set by focuser driver to zero.
Then a focuser new target value is calculated and used.

Similar control is used for the telescope. There offsets are used to command
dithering from the target position.

\subsubsection{Camera acquisition modes}

New modern cameras provide various settings. Some of the combinations of
settings are very good for astronomy, while some are pretty bad.  A desired behaviour
of the system is to allow the user to set only good settings, avoiding the bad ones,
while still leaving an option to set all variables manually.

A desired behaviour of the system will also switch the camera at the end of the
script to a default mode so that the next script will be able to use camera, without
changing any variable.

This functionality is provided by setting the camera mode back to default at the end
of the script. A camera mode file provides various modes. The first is the
default one, which is used to set camera. An example configuration file is
included in the software distribution.

\subsection{Observatory scheduling}

It is widely known that observatory scheduling is not a trivial problem.
Scheduling is known to belong to NP--hard class of problems, which does not make
the problem easier. On top of that, it is not always clear what exactly should
be observed in order to maximise the scientific output of the observatory.

{\it RTS2} currently provides users with three scheduling modes: dispatch
scheduling, queue scheduling and preprogrammed night plan. The advanced,
genetic algorithm scheduling, which should be able to schedule full night runs,
is in the late integration phase. The scheduling modes are described in the
following sections.

\subsubsection{Dispatch scheduling}

Dispatch scheduling is the kind of scheduling used on most, if not all fully
autonomous observatories. Each observable target is assigned a merit function
-- its expected benefits. Dispatch scheduling then calculates merits for all
targets, picks the one with the best merit, and observes it. After this target is
finished, the scheduler recalculates merit functions, and picks a new target. This
approach is discussed by multiple authors, notably\cite{fraser06} and
\cite{tsapras09}. {\it RTS2} provides this mode in the {\it selector} component.

\subsubsection{Queue scheduling}

Queue scheduling schedules targets with the human -- night observer in the
loop. The night observer is presented with a list of possible targets, which
are worth observing. Based on the current conditions, instrument setup and
astronomer preferences, he/she picks a queue and executes its observation. After
the target is finished, a new target is picked from the queue, or the queue is changed by
the observer.

This scheduling is used by current big observatories -- for example by ESO
VLT\cite{chavan97}, CAHA and IAC telescopes.

The observer can fill a queue of targets in the {\it RTS2 executor} component.
There are plans to provide automatic selection from those targets -- so the
observer enters them in the evening, and the system will observe them during
the night at optimal weather and time.

\subsubsection{Night plan}

Another option is to provide a detailed night plan, which will list sequence of
observations, their time and how images should be handled. Although {\it RTS2}
does provide support for night plan, it does not provide any interfaces for
plan creation and its management. The observer must fill a text file and load it
into a database.  Because of the complexity of this operation and lack of support
tools this option is not widely used.

\subsubsection{Genetic Algorithm based scheduling}

None of the scheduling modes discussed above is optimal. Dispatch scheduling
lacks predictability and is short--sighted -- it produces schedules which search
for a local maxima, instead of focusing on long--term gains. Queue scheduling
  and night planning requires non--trivial involvement of the observer.

We proposed and implemented scheduling based on genetics algorithms (GA). The
problem is described in full depth in \cite{pkub08sched}. Here is outline of
this approach.

Observation targets can list various constraints and merits. The algorithm
then searches for the Pareto optimal front\cite{pareto06} -- a set of schedules
which does not violate any constrains and their respective merit functions are
at least comparable to the other best schedules. The search for this front is
performed with NSGA-II\cite{deb00}.

Experimental implementation of GA scheduling is part of the current {\it RTS2}
developer branch. The main benefits of GA scheduling are a simple addition of
new constraints and merits. Ease of its reuse is demonstrated in~\cite{fosterl09}.

The plan is either to provide the observer with Pareto front schedules and let
him/her pick the one he/she likes, or to have system select autonomously during
night the path which will be followed.

\subsection{Weather blocking}
\label{sec:5}

It is very important to close the observatory roof when conditions are hostile for
its normal operations and keeping it closed as long as those conditions
prevail. Our records clearly indicate that having the roof open with bad
weather over it produces significant problems as soon as the wrong roof state
is detected.

The following sections deal with this problem. First the original approach
is described. This is followed by a detailed description of current hardware and
software setup, which is protecting equipment against the elements.

\subsubsection{Original weather reporting}

Keeping track of observatory conditions was originally a job of the dome module. It
was regarded as a single point of failure. Dome code was handled with extreme
care, and dome software was carefully tested after each software upgrade. Each
device which might indicate bad weather sends information about weather state
directly to the dome module. The dome module puts them together, decides if weather
is favourable for the observation, and reacts accordingly - if all conditions
were satisfied, it switched the system to "on" mode. If a single condition was not
met, the dome control module tried to close the roof and switch the system to
"standby" mode.

This solutions has the following problems:

\begin{itemize}

\item dome software testing is time consuming

\item in case of a serious computer hardware problem the dome can be left open

\item every new weather sensor requires modifications to dome control software
and extensive testing before being put in operation

\end{itemize}

\subsubsection{Design of the state based weather handling}

As different new sensors were added to the observatory setup, it becomes clear that
weather reporting deserves special attention. The following requirements were
put for the new algorithm:

\begin{itemize}

\item it must be easy to add new sensors for weather reporting

\item it must be possible to configure the system to include various sensors

\item the system must report bad weather in the event of a sensor failure

\end{itemize}

Apart from those software requirements, additional requirements were put on
the hardware responsible for roof operations:

\begin{itemize}

\item it must run independently from controlling computer

\item it shall be as reliable as possible

\item it shall be as simple as possible

\item it must close the roof on its own when it loses connection to its controlling computer

\item it must report to the computer the states of all sensors connected to it

\end{itemize}

The software and hardware implementation which fulfills all those requiremens is
described in the next section.

\subsubsection{Current weather blocking}

{\it RTS2} centrald and all devices have a state. When changed, the state is
propagated to all connections.

One of the bits in the state represents bad weather. If that bit is set, it
means that the component (in case of device) or whole system (in case of
centrald) reports violation of observing conditions, and therefore asks for
controlled system shutdown.

Centrald also holds a list of devices which are mandatory for observation. If any
of the mandatory devices is not present, or do not reply to centrald requests,
the system is switched to bad weather state. This presents developers with a very
simple way to add a new weather sensor to the system. They can use methods
for weather state manipulation provided in {\it SensorWeather} class.

The ultimate weather protection element is on most systems a programmable logic
controller (PLC). The inputs of this hardware are connected to the various
sensors carring information about the roof state. The outputs of the relay are
connected to the motors responsible for roof operations. The PLC is conservatively
programmed. It is actually harder to open the roof then to close it.

\begin{figure}
\centering
\includegraphics[width=12cm]{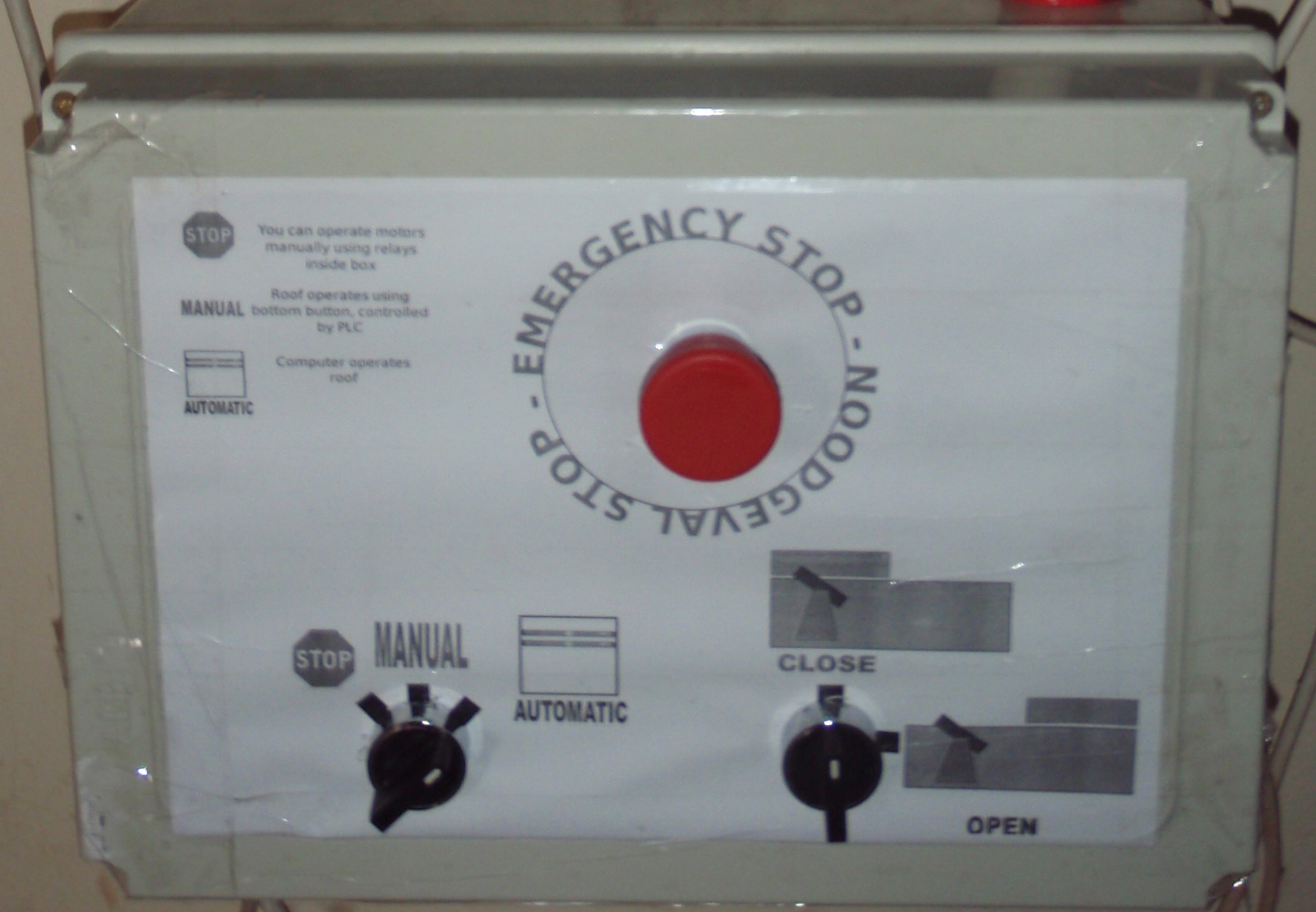}
\caption{Roof control box switches. Switching roof to manual or stop also stops
autonomous observations. Unlike (graphical) user interfaces running on
computers, it is very unlikely that this panel will fail. Watcher telescope,
Boyden Observatory, Republic of South Africa.  }
\label{switches}
\end{figure}

PLC can be controlled manually with switches. For an image of actual roof
control panel please see figure~\ref{switches}.

If the computer would like to open the roof it must send signals every few seconds
to the PLC, to check that it is alive.
The PLC is programed to close the roof if this {\it "I am alive"} signal is not received.

The critical failure points of this setup are: motor failure, failure of power wire
to motors and signals to PLC, PLC failure or power failure.

We do not have capability to make motors redundant. PLCs are designed for harsh
industrial conditions. The PLC program is relatively simple and is very carefully
tested. As its installation requires the presence of the person installing it on
site, it is not sensitive to usual bugs introduced by system upgrades. Power
backup is provided by UPS capable of closing the roof. The UPS state is monitored by
{\it RTS2} and the roof is commanded to close if the remaining UPS uptime or battery
level drops below a predefined values.

PLCs are also commodity value. In the unlikely event that the PLC is damaged or
destroyed, a new can be purchased without significant problems on almost any
place of the planet and installed within a few days from time the failure was
detected.

Source code of the PLC program is available from {\it RTS2} subversion
repository. As was shown by installing it on multiple sites, this setup is
fully replicable and quite modular.

\section{RTS2 installations}

Table \ref{tab:1} provides list of the current {\it RTS2} installations
together with the year they started using {\it RTS2} and current diameter of the
primary optic. The full listing of {\it RTS2} installations is provided on
project web pages.

\begin{table}
\centering
\caption{RTS2 installations}
\label{tab:1}

\begin{tabular}{llrr}
{\bf Name}  &  {\bf Location} & {\bf Year} & {\bf D} \\
\hline
\hline
BART        & Ond\^rejov, Czech Republic     & 2001  & 25cm \\
BOOTES 1    & INTA El Arenossilo, Spain      & 2002  & 30cm \\
BOOTES 2    & La Majora, Spain               & 2003  & 60cm \\
FRAM        & Pierre--Auger South, Argentina & 2004  & 30cm \\
Watcher     & Bloemfotein, South Africa      & 2005  & 40cm \\
BOOTES IR   & OSN, Sierra Nevada, Spain      & 2005  & 60cm \\
LSST CCD testing & BNL, New York, USA        & 2007  & \\
Columbia    & UC, New York, USA              & 2007  & \\
D50         & Ond\^rejov, Czech Republic     & 2008  & 50cm \\
BOOTES 3    & Blenheim, New Zealand          & 2009  & 60cm \\
CAHA 1.23m  & CAHA, Spain                    & 2009  & 1.23m \\
LSST CCD testing & LPNHE Paris, France       & 2009  & \\
\end{tabular}
\end{table}

The system is currently being considered for various new and refurbished
telescopes. Please note that 1.23m at CAHA is currently operated in
semi--automatic mode, and {\it RTS2} is not on controls only during a few
nights. More details about this can be found in \cite{Javier}.

\section{Coding strategy - to restart or to code properly?}

At the beginning of the project, we sometimes employed a "last chance" strategy
where if something fails \footnote{which in our case usually means "produces
core dump"}, the system will notice it, wait for a while and restart the failed
part. Although this looked as very wise approach, it turned out that it hides
potential serious problems and can affect system performance. Moreover, with
errors in certain places it can produce situations when devices will keep
restarting, produce a few lines in the system log, and exit.

Our experience is that the {\it if it fails, let system restart it and hope for
the best} strategy is contra--productive and should not be employed. After
fixing the most important bugs in the code, the programmes run on some setups
for months without need for a single restart. If a problem is encountered, its
root cause is found and fixed. A fixed driver is then installed and started with
the remote ssh access.

The following things are vital for this strategy to work: compile all code
with debugging option, change ulimit for core dumps to unlimited and knowledge
of how to use {\it gdb}\footnote{GDB web site - {\tt http://www.gdb.org}} to
find cause of the core dump. If all of the previous fails, have {\it
valgrind}\footnote {Valgrind web site - {\tt http://www.valgrind.org}} or
a similar memory profiler installed, with knowledge of how to use it to find
memory allocation problems.

Once the coder has managed all those issues, not only will the system operate
smoothly, but he/she also will not be temped to switch to some language with a
garbage collector\footnote{which promises to free developers from memory
allocation problems -- the most probable reason for failure of the C++ code}.
That is not to say that high-level languages are not fine for some jobs -- we do
use Python for GUI and Java/PHP for Web pages. But in our experience, for
low--level, hardware control algorithms, nothing beats properly designed and
coded C/C++ code.

\section{Expected project changes}

Although {\it RTS2} is able to control autonomous observatories, there is still
a lot which deserves more attention, some solutions and some coding. These
items are presented in the following list.

\subsubsection{Image quality monitoring}

The system currently provides some basic quality checks. Those includes results
of a plate solving routine provided by Jibaro \cite{ugarte05} and/or
Astrometry.net\footnote{Astrometry.net web site: {\tt
http://www.astrometry.net}} packages, so observers knows if telescope is
pointing towards the expected position.

The system lacks real time display of various image quality parameters - average
and median values, minimal and maximal values, number of detected objects
(stars) in the image, and so on. Those are calculated and saved to the image
header. Our experience shows that those should be easily accessible in real--time
displays, as well as in tool for navigation through the image archive.

\subsubsection{Image processing}

Image processing can be configured as an external script, providing relative
photometry and other services. We would like to better integrate it with the
system by:

\begin{itemize}

\item adding calibration image database, and providing tools to maintain and
use this database

\item providing observations--to--paper solutions for predefined well understood
problems -- among others observatorion of planetary transit occultations and
micro--lensing events

\end{itemize}

\subsubsection{Better user interfaces}

The system is primarily controlled through an ncurses based interface. While this is
sufficient for experienced users, the interface is not well suited for
occasional users.

Development of both Web and X-Windows Graphical User Interfaces is currently in
initial phases. Plans also call for addition of applets for GNOME and other
desktops environments, so we will be able to track telescope operations
world--wide from a single desktop.

\subsubsection{Network management}

The System currently lacks a central management console. Observatories are usually
managed and monitored through ssh connections. While this approach is feasible,
it quickly grows beyond the capabilities of a single maintainer.

There is a need for application, which will:

\begin{itemize}

\item show states of individual observatories in the network

\item list unresolved problems of individual telescopes, keeping track
of actions to fix the problem, and the results of those actions

\item enable scheduling of the whole network

\item synthesise results obtained by network members

\end{itemize}

It would be nice to have those features backed by simple, low-level, Internet
interface. There is work in progress on an XML-RPC interface which will do
just this.

\subsubsection{Archive access}

{\it RTS2} keeps all important information in a database. Information about
executed observations, and images acquired, together with basic image
characteristics (which includes WCS coordinates, fitted by Jibaro and/or
Astrometry.net) is recorded for quick retrieval.

Current archive access is provided primarily through console based utilities.
Without doubt in this is a bit too old fashioned in the graphical user
interfaces age.

There were PHP scripts for Web based archive access, which even included
cut-out service. Unfortunately they were not very well designed. As the project
matures and introduces changes to the database, those scripts cease to
function, to the point when it was ruled to be too expansive to make them work
again - as is the usual case with simple small fast PHP and other scripts.

Up to now multiple attempts to provide better solutions were made. Yet till
now each of them has failed to deliver usable results.

Current attempts include a Google Web Toolkit XML-RPC backed application. We
have reasonable hopes that this will deliver promised results, although not in
a short timescale.

\section{Conclusions}
\label{sec:5}
This article presented an open-source system for robotic observatory control. It
provides an overview of rationales for its development, its composition and its
main features. It then focused on a list of recent improvements. The article
concludes with lists of development items still left on the agenda.

\vskip 1cm
{\it Acknowledgement}
The author would like to acknowledge generous financial support provided by
Spanish {\it Programa de Ayudas FPI del Ministerio de Ciencia e Innovación
(Subprograma FPI-MICINN)} and European {\it Fondo Social Europeo}. Work on {\it
RTS2} was supported, influenced and encouraged by numerous people, whose list
would be too large for this article. Persons from this list which according to
author deserve to be mention explicitly are Martin Jel\'inek, Alberto
Castro-Tirado, Antonio de Ugarte Postigo, Ronan Cunniffe, Michael Prouza,
Ren\'e Hudec, Victor Reglero and Beatriz S\'anchez F\'elix. The article was
carefully reviewed by two anonymous referees, whose suggestions significantly
improved it. Stephen Bailey made final gramatical improvements to the article.

\end{document}